\documentclass{pos}
\newcommand{\gsim}{\raisebox{-0.07cm   }
{$\, \stackrel{>}{{\scriptstyle\sim}}\, $}}
\newcommand{\ep}{\varepsilon}
\newcommand{\N}{\nonumber}

\title{{\footnotesize DESY 08-186}\\
Heavy flavor operator matrix elements at 
{${O(a_s^3)}$}}

\ShortTitle{Heavy flavor operator matrix elements at 
$O(a_s^3)$}

\author{Isabella Bierenbaum\thanks{Present address: Instituto de F\'{i}sica 
Corpuscular, CSIC-Universitat de Val\`{e}ncia, Apartado de Correros 22085, 
E-46071 Valencia, Spain. Partially supported by the 
Ministerio de
    Ciencia e Innovaci\'on under Grant No. FPA2007-60323, by CPAN (Grant
    No. CSD2007- 00042), and by the European Commission MRTN FLAVIAnet under
    Contract No. MRTN-CT-2006-035482.}, 
~~Johannes Bl\"umlein,~~\speaker{Sebastian 
Klein}
        \thanks{Supported in part by Studienstiftung des 
                Deutschen Volkes.} \\
        Deutsches Elektronen--Synchrotron, DESY \\
        Platanenallee 6, D--15738 Zeuthen, Germany \\
        E-mail: 
\email{Isabella.Bierenbaum@ific.uv.es, Johannes.Bluemlein@desy.de, Sebastian.Klein@desy.de}}
\abstract{
\noindent  
 The heavy quark effects in deep--inelastic scattering 
 in the asymptotic regime $Q^2 \gg m^2$ can be described 
 by heavy flavor operator matrix elements. Complete analytic expressions 
 for these objects are currently known to ${\sf NLO}$. We present 
 first results for fixed moments at ${\sf NNLO}$. This involves 
 a recalculation of fixed moments of the corresponding ${\sf NNLO}$ anomalous
 dimensions, which we thereby confirm.
 }
\FullConference{8th Conference Quark Confinement and the Hadron Spectrum\\
                September 1-6, 2008\\
                Mainz. Germany}

\begin{document}
%
%
%
\section{Introduction}
\noindent
The double--differential cross section for  unpolarized deep inelastic
scattering via single photon exchange can be expressed in terms of two
structure functions $F_{2,L}(x,Q^2)$. Especially in the region of smaller
values of $x$, these structure functions contain large
$c\overline{c}$--contributions of up to 20-40~\%, denoted by
$F_{2,L}^{c\overline{c}}(x,Q^2)$
\footnote{We consider extrinsic heavy flavor production only.}. The
perturbative heavy flavor Wilson coefficients $H_{2,L}(z,Q^2,m^2)$
corresponding to these structure functions are known at ${\sf NLO}$
semi--analytically in $z$--space, \cite{HEAV1}, with a fast numerical
implementation in Mellin--$N$ space given in \cite{HEAV2}. Due to the size of
the heavy flavor corrections, it is necessary to extend the description of
these contributions to $O(a_s^3)$, $a_s:=\alpha_s/(4\pi)$, and thus to the
same level which has been reached for the massless Wilson coefficients,
\cite{MVV}. This will allow for a more precise determination of parton
distribution functions and the ${\sf QCD}$--scale $\Lambda$.  A calculation of
these quantities in the whole kinematic range at ${\sf NNLO}$ seems to be out
of reach at present. However, in the limit of large virtualities $Q^2$, $Q^2\:
\gsim \:10\:m_c^2$ in the case of $F_2^{c\bar{c}}(x,Q^2)$, one observes that
$F_{2,L}^{c\bar{c}}(x,Q^2)$ are very well described by their asymptotic
expressions, \cite{BUZA}, neglecting power corrections in $m^2/Q^2$.  In this
kinematic range, one can calculate the heavy flavor Wilson coefficients
analytically. This has been done for $F^{c\bar{c}}_2(x,Q^2)$ to 2--loop order
in \cite{BUZA,BBK1} and for $F^{c\bar{c}}_L(x,Q^2)$ to 3--loop order in
\cite{BFNK}. In the latter case, the asymptotic result becomes valid only at
much higher values of $Q^2$. As first steps towards the 3--loop calculation at
asymptotic scales for $F_2^{c\bar{c}}(x,Q^2)$ we calculated the O($\ep$) terms
of the 2-loop heavy operator matrix elements (OMEs),
\cite{UnPolOeps,KrakLLProc}, which contribute to the asymptotic 3--loop heavy
flavor Wilson coefficients via renormalization. In the present paper, we
report on new results concerning moments of the heavy OMEs $A_{Q(q)q}^{\sf
  PS}$, $A_{qq,Q}^{\sf NS,+}$ and $A_{gq,Q}$ at $3$--loops. In doing so, the
fermionic terms of the even moments $N=2...12$ of the corresponding ${\sf
  NNLO}$ anomalous dimensions given in \cite{ANDIMNSS} are confirmed in an
independent calculation.
%
%
%
\section{Heavy Flavor Operator Matrix Elements}
\noindent
In the following, we consider massive OMEs of the 
flavor--decomposed twist--2 operators between partonic states
\begin{eqnarray}
  A_{ki}^{\sf S, NS}\Bigl(\frac{m^2}{\mu^2},N\Bigr)  = \langle 
  i|O_k^{\sf S, NS}|i\rangle_H
  = \delta_{k,i} + \sum_{l=1}^{\infty} a_s^l 
     A_{ki}^{{\sf S, NS},(l)}\Bigl(\frac{m^2}{\mu^2},N\Bigr)~.
       \label{OMEs}
\end{eqnarray}
Here, ${\sf S}$ and ${\sf NS}$ are the singlet and non--singlet contributions,
respectively, $i$ denotes the outer on--shell particle ($i=q,g$) and $O_k$
stands for the quarkonic ($k=q$) or gluonic ($k=g$) operator emerging in the
light--cone expansion. The subscript $H$ indicates that we require the
presence of heavy quarks of one type with mass $m$, while $\mu$ is the
renormalization scale. We work in Mellin--space, with the Mellin--variable
denoted by $N$.  The logarithmic terms in $m^2/\mu^2$ are completely
determined by renormalization and contain contributions of the anomalous
dimensions of the twist--2 operators.  Thus at ${\sf NNLO}$ the fermionic
parts of the 3-loop anomalous dimensions calculated in Refs. \cite{ANDIMNSS}
appear. All pole terms provide a check on our calculation. The single pole
term allows for a first independent calculation of the terms $\propto T_F$ in
the 3-loop anomalous dimensions.\\ As outlined in Ref.~\cite{BUZA}, in the
limit $Q^2 \gg m^2$ one applies the massive renormalization group equation to
obtain the factorization of the heavy flavor Wilson coefficients into a
Mellin--convolution of the light flavor Wilson coefficients with the
corresponding heavy OMEs. The light flavor Wilson coefficients are known up to
three loop order \cite{MVV} and carry all the process dependence, whereas the
heavy quark OMEs are universal, process--independent objects and contain all
mass corrections to $H_{2,L}$ up to terms proportional to $(m^2/Q^2)^a,~a \ge 1$. \\ A
related application of the heavy OMEs is given when using a variable flavor
number scheme to describe parton densities including massive partons. The OMEs
are then the transition functions when going from $n_f$ to $n_f+1$ flavors and
thus one may define parton densities for massive quarks, see
e.g. Ref. \cite{Buza:1996wv}. This is of particular interest for heavy quark
induced processes at the LHC, such as $c\overline{s}\rightarrow W^+$ at large
enough scales $Q^2$.  In this context, one may show as well that various sum
rules follow from momentum conservation, \cite{Buza:1996wv}, e.g:
\begin{eqnarray}
  A_{qq,Q}^{ \sf NS,+}\Bigl|_{N=2}
 +A_{Q(q)q}^{ \sf PS}\Bigl|_{N=2}
 +A_{gq,Q}\Bigl|_{N=2}
 &=& 0 ~. \label{sumrule}
\end{eqnarray}
Note, that in Eq. (\ref{sumrule}) two different ${\sf PS}$ contributions can be
distinguished and we adopt the notation $A_{Q(q)q}^{\sf PS}:=A_{Qq}^{\sf
  PS}+A_{qq,Q}^{\sf PS}$, where for the $\{Qq\}$-- term the operator 
couples to the heavy quark and for the $\{qq,Q\}$--term 
to a light quark. 
%
%
%
\section{Renormalization}
\noindent
We work in Feynman--gauge and use dimensional regularization in $D=4+\ep$
dimensions, applying the $\overline{\rm{MS}}$--scheme, if not stated
otherwise. Renormalization proceeds in four steps, which we will briefly
sketch here and refer to 
\cite{BUZA,BBK1,UnPolOeps,KrakLLProc} for more details.  Mass renormalization 
is performed in
the on--shell scheme \cite{MASS2}, whereas for charge renormalization we use
the $\overline{\rm{MS}}$--scheme. The remaining singularities are of the
ultraviolet and collinear type. The former are renormalized via the operator
$Z$--factors, whereas the latter are removed via mass factorization through
the transition functions $\Gamma$.
After coupling-- and mass renormalization, the renormalized heavy flavor OMEs 
are
then obtained via
\begin{equation}
 A=Z^{-1} \hat{A}  \Gamma^{-1}~, \label{GenRen}
\end{equation}
where quantities with a hat are unrenormalized. Note, that in the singlet case
operator mixing occurs and hence Eq. (\ref{GenRen}) should be read as a matrix
equation, contrary to the ${\sf NS}$--case.  The $Z$-- and $\Gamma$--factors
can be expressed in terms of the anomalous dimensions of the twist--$2$
operators in all orders in the strong coupling constant $a_s$,
cf. \cite{UnPolOeps} up to $O(a_s^3)$. For this purpose, we adopt the
convention
\begin{eqnarray}
 { \gamma} = \mu \partial \ln { Z(\mu)}/ \partial \mu~.
\label{gamzet}
\end{eqnarray}
From Eqs. (\ref{GenRen},\ref{gamzet}) one can then infer that for operator
renormalization and mass factorization at $O(a_s^3)$, the anomalous dimensions
up to ${\sf NNLO}$, \cite{ANDIMNSS}, together with the $1$--loop heavy flavor
OMEs up to $O(\ep^2)$ and the $2$--loop heavy OMEs up to $O(\ep)$ are
needed. Higher orders in $\ep$ enter since they multiply $Z-$ and
$\Gamma$--factors containing poles in $\ep$. This has been worked out in some
detail in Ref. \cite{UnPolOeps}, where we presented the $O(\ep)$ terms
$\overline{a}_{Qg}^{(2)}$, $\overline{a}_{qq,Q}^{(2), {\sf NS}}$ and
$\overline{a}_{Qq}^{(2){\sf PS}}$ in the unpolarized case. The terms
$\overline{a}_{gg,Q}^{(2)}$ and $\overline{a}_{gq,Q}^{(2)}$ were given in
Refs. \cite{KrakLLProc}.  Thus all terms needed for the renormalization at
$3$--loops in the unpolarized case are known by now.
%
%
%
\section{Moments of the Massive OMEs at three Loops}
\noindent
The heavy flavor OMEs at $O(a_s^3)$ are given by $3$--loop self--energy type
diagrams, which contain a local operator insertion. The external massless
particles are on--shell. The heavy quark mass sets the scale and the spin of
the local operator is given by the Mellin--variable $N$. The steps for the
calculation are the following: We use {\sf QGRAF}, \cite{Nogueira}, for the
generation of diagrams.  Approximately $100$ diagrams contribute for each case
at hand.  For the calculation of the color factors we use \cite{COLORF}.
After undoing the contraction of the operators with the light--like vector
$\Delta$,~~($\Delta^2=0$), the diagrams are genuinely given as tensor
integrals. Applying a projector then provides the results for the diagrams for
the specific (even) Mellin moment under consideration.  The diagrams are
further translated into a form, which is suitable for the program ${\sf
  MATAD}$ \cite{MATAD}, through which the expansion in $\ep$ is performed and
the corresponding massive three--loop tadpole--type diagrams are
calculated. We have implemented all these steps into a {\sf FORM}--program,
cf.~\cite{FORM}, and use {\sf TFORM}, \cite{TFORM}, for parts of the
calculation. We checked our procedures against various complete two--loop
results and certain scalar $3$--loop integrals and found full agreement.
%
%
%
\section{Results}
 \noindent
 Applying Eq. (\ref{GenRen}), one can predict the pole structure of the
 unrenormalized results and thus the logarithmic terms of the renormalized
 OMEs. Since we consider only terms involving at least one heavy
 quark, we adopt the following notation for the anomalous dimensions
 \begin{eqnarray}
  \hat{\gamma}\equiv \gamma(n_f+1)-\gamma(n_f)~.  \label{hatpres}
 \end{eqnarray}
 As an example, we show the structure of the renormalized result in the 
 ${\sf PS}$ case, where all renormalization constants are taken at 
 $n_f$ flavors.
 \begin{eqnarray}
   {A_{Q(q)q}^{(3),{\sf PS}}}&=&
      \frac{{\hat{\gamma}_{qg}^{(0)}}{\gamma_{gq}^{(0)}}}{48}
                  \Biggl\{
                         {\gamma_{gg}^{(0)}}
                        -{\gamma_{qq}^{(0)}}
                        +4({n_f}+1){\beta_{0,Q}}
                        +6{\beta_0}
                  \Biggr\}
              {\ln^3 \Bigl(\frac{m^2}{\mu^2}\Bigr)}
  +               \Biggl\{
                         \frac{{\hat{\gamma}_{qq}^{(1),{\sf PS}}}}{2}
                               \Bigl(
                                 ({n_f}+1){\beta_{0,Q}}
                                -{\beta_0}
                               \Bigr)
\N\\ &&
                        +\frac{{\hat{\gamma}_{qg}^{(0)}}}{8}
                               \Bigl(
                                 ({n_f}+1){\hat{\gamma}_{gq}^{(1)}}
                                -{\gamma_{gq}^{(1)}}
                               \Bigr)
                        -\frac{{\gamma_{gq}^{(0)}}{\hat{\gamma}_{qg}^{(1)}}}{8}
                  \Biggr\}
              {\ln^2 \Bigl(\frac{m^2}{\mu^2}\Bigr)}
  +               \Biggl\{
                       \frac{\hat{\gamma}_{qq}^{(2),{\sf PS}}}{2}
                        -2{a_{Qq}^{(2),{\sf PS}}}{\beta_0}
                        -\frac{{\gamma_{gq}^{(0)}}}{2}{a_{Qg}^{(2)}}
\N\\ &&
                        -\zeta_2\frac{{\hat{\gamma}_{qg}^{(0)}}
                                      {\gamma_{gq}^{(0)}}}{16}
                          \Bigl(
                                 {\gamma_{gg}^{(0)}}
                                -{\gamma_{qq}^{(0)}}
                                +4({n_f}+1){\beta_{0,Q}}
                                +6{\beta_0}
                          \Bigr)
                        +\frac{{n_f}+1}{2}{\hat{\gamma}_{qg}^{(0)}}{a_{gq,Q}^{(2)}}
                  \Biggr\}
              {\ln \Bigl(\frac{m^2}{\mu^2}\Bigr)}
  +{\gamma_{gq}^{(0)}}{\overline{a}_{Qg}^{(2)}}
\N\\ &&
   +\zeta_3\frac{{\gamma_{gq}^{(0)}}{\hat{\gamma}_{qg}^{(0)}}}{48}
                  \Bigl(
                         {\gamma_{gg}^{(0)}}
                        -{\gamma_{qq}^{(0)}}
                        +4{n_f}{\beta_{0,Q}}
                        +6{\beta_0}
                  \Bigr)
  +\frac{\zeta_2}{16}
                  \Bigl(
                        -4{n_f}{\beta_{0,Q}}{\hat{\gamma}_{qq}^{(1),{\sf PS}}}
                        +{\hat{\gamma}_{qg}^{(0)}}{\gamma_{gq}^{(1)}}
                  \Bigr)
\N\\ &&
  +4({\beta_0}+{\beta_{0,Q}}){\overline{a}_{Qq}^{(2),{\sf PS}}}
  +C_F
                 \Bigl(
                       -(4+\frac{3}{4}\zeta_2)
                            {\hat{\gamma}_{qg}^{(0)}}{\gamma_{gq}^{(0)}}
                      -4{\hat{\gamma}_{qq}^{(1),{\sf PS}}}
                      +12{a_{Qq}^{(2),{\sf PS}}}
                 \Bigr)
\N\\ &&
  -({n_f}+1){\hat{\gamma}_{qg}^{(0)}}{\overline{a}_{gq,Q}^{(2)}}
  +{a_{Q(q)q}^{(3),{\sf PS}}}~. \label{AQqq3PSren}
  \end{eqnarray}
 Here, the terms $a_{ij}^{(2)}$ denote the constant terms in $\ep$ of the
 $2$--loop OMEs $\hat{A}_{ij}$ and $\beta_i$ are the expansion coefficients of
 the $\beta$--function. The subscript $Q$ refers to contributions due to heavy
 quarks only, cf. \cite{BUZA,BBK1,UnPolOeps}, and $\zeta_i$ is the Riemann
 $\zeta$--function at values $i$.  All quantities in Eq. (\ref{AQqq3PSren})
 are known for general values of $N$, except for $a_{Q(q)q}^{(3)}$, which is 
the genuine
 $3$--loop contribution and remains to be calculated. Note, that it is not
 possible to factor out $n_f+1$, not even in the triple pole term. This is due
 to the interplay of the prescription for coupling constant renormalization we
 have adopted, cf. \cite{BUZA,UnPolOeps}, and the fact that the transition
 functions $\Gamma$ apply to sub--graphs containing massless lines only. \\
 \noindent
  We have calculated the OMEs $A_{Q(q)q}^{{\sf PS}, (3)}$, $A_{qq,Q}^{{\sf
      NS,+}, (3)}$ and $A_{gq,Q}^{(3)}$ for $N=2...12$, using {\sf MATAD}.
  All pole terms agree with the general structure derived from
  renormalization.  As an example, we show the constant term after
  renormalization of the ${\sf PS}$--OME for $N=2$
  \begin{eqnarray}
   &&A_{Q(q)q}^{(3),{\sf PS}}\Biggl|^{N=2}_{\mu^2=m^2}=
        \Bigl(
               \frac{2048}{81}\zeta_3
              -\frac{86480}{2187}
        \Bigr)C_FT_F^2n_f
       +\Bigl(
              -\frac{3584}{81}\zeta_3
              +\frac{53144}{2187}
        \Bigr)C_FT_F^2
       +\Bigl( 
               \frac{64}{9}B_4
              -\frac{128}{5}\zeta_2^2
\N\\ &&
              +\frac{1280}{27}\zeta_3
              +\frac{830}{2187}
        \Bigr)C_FC_AT_F
       +\Bigl(
              -\frac{128}{9}B_4
              +\frac{128}{5}\zeta_2^2
              -\frac{9536}{81}\zeta_3
              +\frac{95638}{729}
        \Bigr)C_F^2T_F ~. \label{AQqq3PS2}
  \end{eqnarray}
 The moments we obtain for $N=10,~12$ for the corresponding $3$--loop
 anomalous dimensions are shown in Table \ref{table:resultsandim1012}.  These
 results, as well as the moments we obtain for lower values of $N$, agree with
 the results of Refs. \cite{ANDIMNSS}. A check for the constant terms is
 provided by the sum rule in Eq. (\ref{sumrule}), which is obeyed.  
Additionally, we find for all moments that the terms
 proportional to $\zeta_2$ disappear after renormalization, which is a
 general observation made in many $D \rightarrow 4$ calculations. The term 
$B_4$ in
 Eq. (\ref{AQqq3PS2}) is given by
 \begin{eqnarray}
  {\sf B_4}&=&-4\zeta_2\ln^2 2 +\frac{2}{3}\ln^4 2 -\frac{13}{2}\zeta_4
             +16 {\sf Li}_4\Bigl(\frac{1}{2}\Bigr)~, \label{B4}
  \end{eqnarray}
  and appears in all OMEs we calculated. Since it does not appear in
  the light--flavor Wilson coefficients, cf. \cite{MVV}, 
  it occurs as a genuine mass effect.
%
{\small{
\begin{table*}[htb]
\caption{Results for $N=10,~12$ for the $3$--loop anomalous dimensions}
\label{table:resultsandim1012}
\begin{center}
  \includegraphics[height=13.92cm,angle=90]{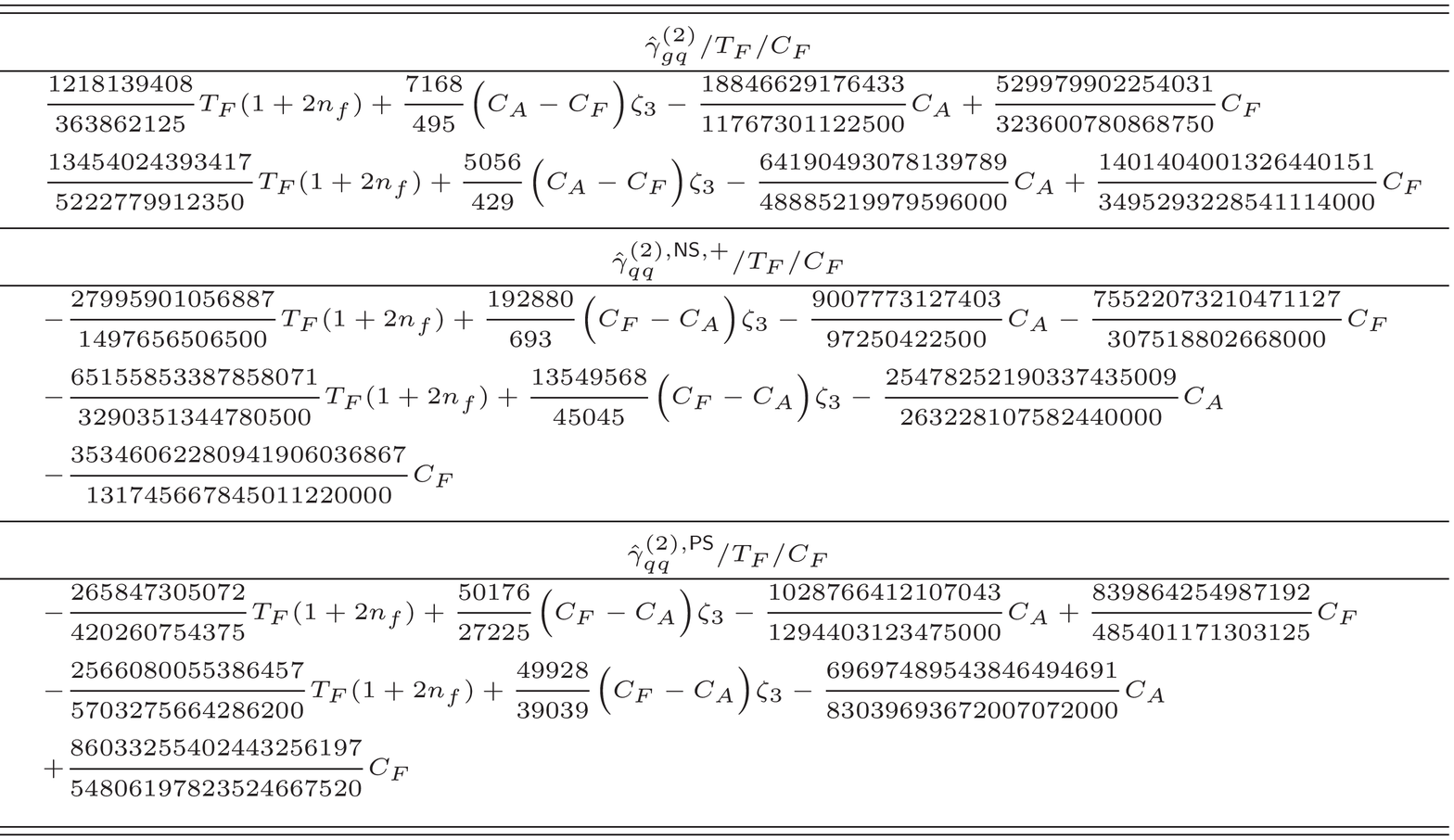}
\end{center}
\end{table*}
\renewcommand{\arraystretch}{1.0} 
}}
%
%
%
\section{Conclusions and Outlook}
\noindent
We calculated the heavy OMEs $A_{qq,Q}^{(3),{\sf NS,+}}$,
$A_{Qq}^{(3),{\sf PS}}$ and $A_{gq,Q}^{(3)}$ for even Mellin--moments
$N=2...12$ using {\sf MATAD} and showed first results. This confirms for the
first time in an independent calculation the moments of the fermionic parts of 
the
corresponding $3$--loop anomalous dimensions, \cite{ANDIMNSS}. We expect
results for the remaining terms $A_{gg,Q}^{(3)}$ and $A_{Qg}^{(3)}$ 
in the near future, thus enabling us to calculate fixed moments of
the heavy flavor Wilson coefficients in the asymptotic limit $Q^2\gg m^2$.

\vspace{5mm}\noindent
{\bf Acknowledgments.}~~
The work was supported in part by SFB--TR--9: Computergest\"utzte
                Theoretische Teilchenphysik and EU-network HEPTOOLS.
We would like to thank M.~Steinhauser 
and J. Vermaseren for useful discussions and M. Steinhauser for a
{\tt FORM 3.0} compatible form of the code {\tt MATAD}. 
%
%
%

\end{document}